\documentclass[12pt]{article}
\usepackage{pdproc,epsfig} 

\begin{document}

\renewcommand{\thefootnote}{\alph{footnote}}
  
\title{GAMMA-RAY BURSTS AND THE SOCIOLOGY OF SCIENCE}

\author{ALVARO DE R\'UJULA}

\address{ CERN\\
 1211 Geneva 23, Switzerland\\
 {\rm E-mail: alvaro.derujula@cern.ch}}

\abstract{I discuss what we have learned about Gamma-Ray Bursts (GRBs)
by studying their {\it afterglows}, and how these are interpreted in the
generally-accepted {\it fireball} model of GRBs, as well as in the 
generally-unaccepted {\it cannonball} model of the same phenomena.
The interpretation of GRBs is a  
good example around which to frame a discussion of the different 
approaches to science found in various fields, such as
high-energy physics (HEP), high-energy
astrophysics, or even the deciphering of ancient languages. 
I use this example to draw conclusions on {\it post-academic}
science, and on the current status of European HEP.}
   
\normalsize\baselineskip=15pt

\section{Motivation}
Why would one give a talk on the sociology of science at a meeting
on {\it Un altro modo di guardare il cielo}, or more generally on
neutrino physics, the meeting
definitely {\bf not} being centred on sociology? My excuse is that, when
talking about GRBs, I have always detected a very considerable interest
on their {\it sociology}: an extremely negative reaction to my views
on the subject in meetings on GRBs, and an extremely surprised
and curiosity-driven interest from HEP or general audiences.

A talk or article such as this one may seem rather unusual, but it is not. 
A precedent with quite similar
content and conclusions has been written by Charles Dermer\cite{Der}.

\section{Introduction}

Some three times a day, on average,
gamma-ray bursts (GRBs) reach the upper atmosphere from
isotropically distributed sky locations.
Much of their energy is in photons of a few hundred keV,
with a total fluence of $10^{-6}$ erg s$^{-1}$, give or take a
couple of orders of magnitude. GRB durations range from tens 
of milliseconds to hundreds of seconds, with varied time
structures generally consisting of fast rising and declining,
isolated or partially-superimposed pulses. The distribution of 
GRB durations is bimodal, with a trough at $\sim 2$ s, separating 
``short'' from ``long'' GRBs.

\begin{figure}
\vspace*{-160pt}
%\leftline{\hfill\vbox{\hrule width 5cm height0.001pt}\hfill}
%\hspace{-.5cm}
\epsfig{figure=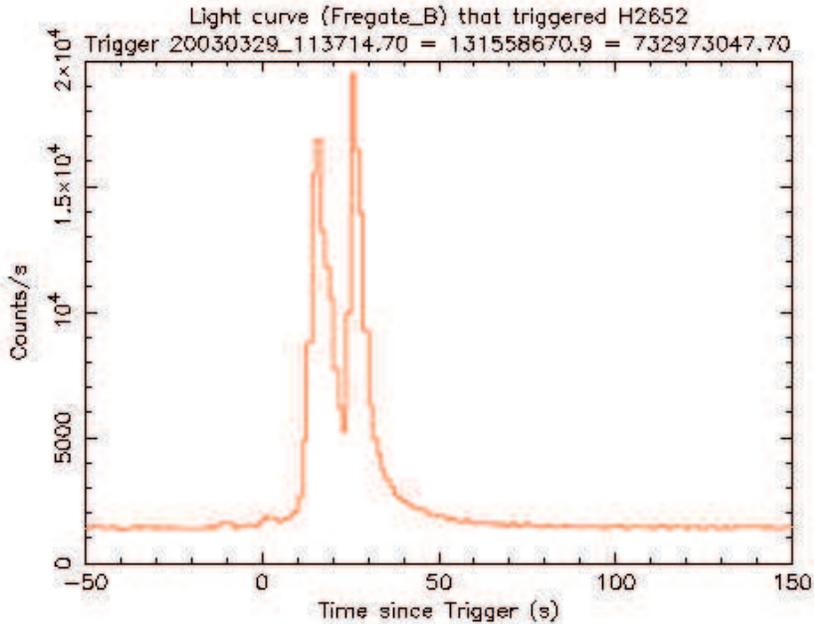,width=14.cm}
\vspace*{-2.2truein}		
%\leftline{\hfill\vbox{\hrule width 5cm height0.001pt}\hfill}
\caption{The $\gamma$-ray counting rate of GRB 030329, observed with the
HETE satellite by Vanderspek et al. The figure is from
http://space.mit.edu/HETE/Bursts/GRB030329/.}
\label{fig:0}
\end{figure}

The example of the $\gamma$-ray counting rate of GRB 030329
is given in Fig.~\ref{fig:0}. This burst, originating at
$z=0.1685$, is the second-closest GRB of known distance.
It is ``long'' and it has a relatively simple structure,
with just two similar ``pulses''.

It has become a custom to refer to Gamma-Ray Bursts as
{\it the biggest explosions since the big bang}.  I
contend that this is bad propaganda: the big bang was not an
explosion in any conventional sense, and GRBs are a highly-beamed
but relatively-small fraction of the energetic budget of
a supernova (SN) explosion. For not-very updated reviews of these
views, and to trace the many references I cannot include, see Dar\cite{Dar} 
and De R\'ujula\cite{ADR} and... references therein. 

We call our unconventional theory of GRBs ``the CannonBall (CB)'' 
model. The conventional models are collectively dubbed the 
``fireball'' model\footnote{These models
are ``standard'' in that the analysis of observations  is
invariably phrased in their language. They are not  ``standard''
in the same sense as the standard model of particle physics.}. 
%which is consistent, univocal, unchallenged by observations and better 
%supported by data than by fans.}. 
In the
available space, I cannot give justice to either model, so I will describe
them rather superficially and compare them in detail only in one particular
but not unusual case, concerning the analysis of the ``radio afterglow''
of GRB 991208.

\section{Science}

\subsection{GRB afterglows}

Our information about the once totally mysterious gamma-ray bursts 
increased spectacularly in the past few years.
The rapid directional localization of GRBs by the
satellites BeppoSAX, Rossi 
and by the Inter-Planetary Network of spacecrafts led to 
a flurry of progress. The crucial discovery\cite{Costa} was
the existence of  ``afterglows'' (AGs) of  long-duration
GRBs: not surprisingly, a GRB ``event'' does not end as the $\gamma$-ray
flux becomes undetectably small. The source continues to emit light
at all smaller observable frequencies, ranging from X-rays to radio waves,
and to be observable for months, or even years. The fact that these
remaining emissions can be very well localized in the sky has led
to the discovery of the GRBs' host galaxies\cite{Sahu}; 
the measurement of their redshifts\cite{Metzger} that verified their
cosmological  origin; the
identification of their birthplaces ---mainly star formation regions in
normal galaxies---  and the first evidence for a possible association 
(in time and location)
between GRBs and supernova explosions\cite{Galama}:
that of GRB 980425 and the supernova SN1998bw.

\begin{figure}
\vspace*{-110pt}
%\leftline{\hfill\vbox{\hrule width 5cm height0.001pt}\hfill}
\hspace{-.5cm}
\mbox{
\epsfig{figure=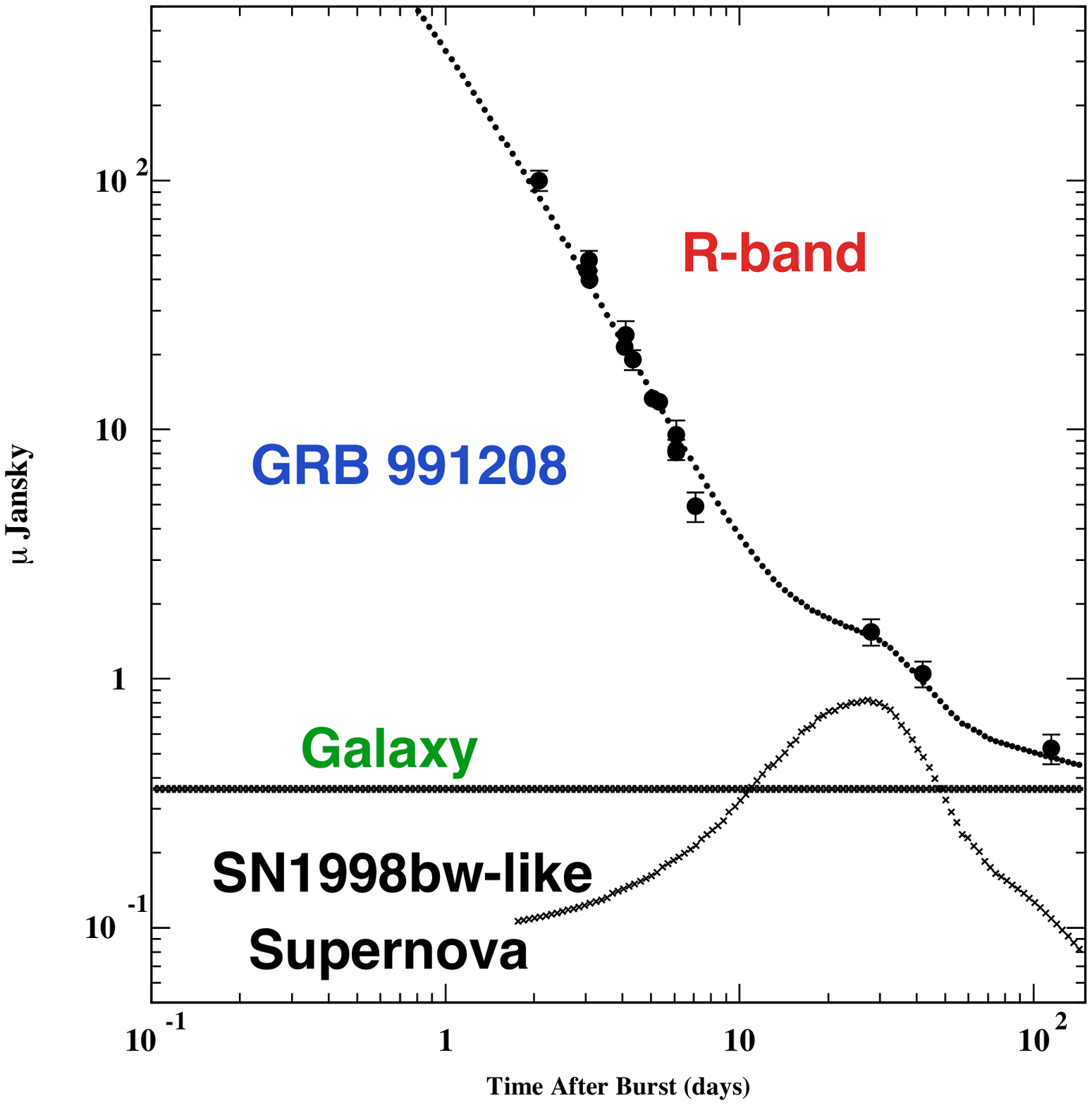,width=8.cm}
\epsfig{figure=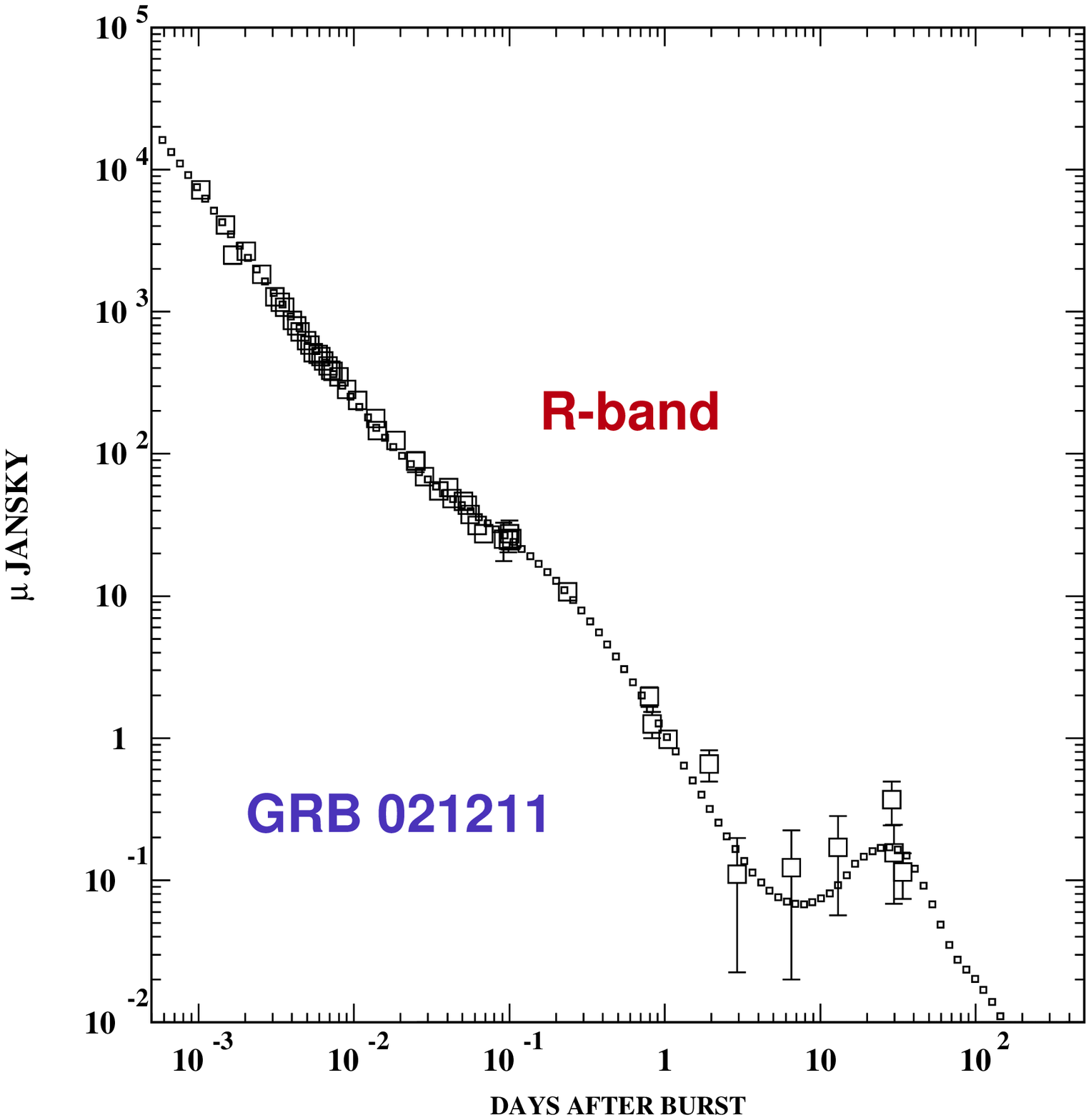,width=8.cm}}
\vspace*{-0.4truein}		
%\leftline{\hfill\vbox{\hrule width 5cm height0.001pt}\hfill}
\caption{Left: the R-Band AG of GRB 990128, showing the various
contributions of a CB-model fit. Right: The R-Band AG of GRB 021211,
in this case the galaxy's fitted contribution is subtracted.
A $\mu$-Jansky is a fluence of $10^{-29}$ erg s$^{-1}$ cm$^{-2}$ Hz$^{-1}$. }
\label{fig:1}
\end{figure}

Two examples of optical ``R-band'' AGs and their CB-model fits are given in 
Fig.~\ref{fig:1}, showing the time evolution of their fluence (measured
energy per unit time, surface and frequency interval).
In the case of GRB 991208, 
%at redshift $z=0.706$,
three contributions
to the observations are shown and added: the unresolved host galaxy,
the ``true'' GRBs AG, and a supernova identical to SN1998bw, but
``transported'' to the redshift of this particular GRB\footnote{The fact
that the progenitors of long GRBs are core-collapse SNe, long advocated
in the CB model, has received spectacular support after this talk was
delivered\cite{ourSN,theirSN}.}. In the case of GRB 021211, the fitted host
galaxy's contribution is subtracted, and the SN contribution to the
AG is also observable.

\subsection{The afterglows of fireballs and firecones}

In the fireball model, reviewed, for instance, by Piran\cite{Piran} 
and Meszaros\cite{Meszaros},
 both the $\gamma$-rays and the AG of a GRB are made by
synchrotron radiation in backward- and forward-moving shocks, which
are produced as relativistically expanding shells collide with each other
and with the interstellar medium (ISM). These shells are made of $e^+ e^-$
pairs, with a small admixture of protons (a ``baryon-load'') fine-tuned
not to quench the observed radiation. 

The possibility that the fire-``ball''
ejecta may not be spherically distributed has been repeatedly
studied in the literature. In the fireball model this was not done
in detail prior to the influential papers by Rhoads\cite{Rhoads},
who predicted abrupt {\it breaks} in the power-law 
temporal behaviour of the AG light curves.
%With the advent of GRB 990123, with its record equivalent spherical 
%energy (see Table I) and an AG light curve through which it is
%possible to draw a broken power law (e.g. Figs. 1--4 of Holland et al.~2000a)
%the fireball advocates (see, e.g. Frail et al.~2001) adopted the 
%arguments in favour of collimated GRBs (e.g. in the case of
%GRBs from quasars, Brainerd 1992; in the case of a funnel in
%an explosion, Meszaros \&
%Rees 1992; and in the case of jets in gravitational collapses,
%Shaviv \& Dar 1995, Dar 1997, Dar 1998a; Dar \& Plaga 1999, DD2000a and
%references therein). 
So did fireballs evolve into ``collimated fireballs'',
``firecones'' or ``conical fireball jets'', while maintaining the ``fire'' 
lineage.

As in the cases of Fig.~\ref{fig:1}, abrupt breaks are not observed. 
Yet,  observers fit 
AG light curves to extract a {\it break time} $t_b$ from 
``phenomenological'' formulae, such as:
\begin{equation}
F_\nu = {2\,F_\nu^b \over 
\left[ (t/t_b)^{\alpha_1\,s}+(t/t_b)^{\alpha_2\,s}\right]^{1\over s}}\, ,
\label{pheno}
\end{equation}
which interpolate between two power laws with a tunable ``abruptness'' 
$ s$, often set to $s=1$. 
The values of $ t_b$ extracted from such arbitrary
expressions have no clear meaning. Even when applied
to the same GRB's AG, a fit of the above form may be problematic.
In the case of GRB 020813, for instance, Covino et al.~extract
$t_b=0.59\pm 0.03$ days from data in the interval between
3 hours and 4 days\cite{Covino}, while Li et al.~find $t_b=0.13\pm 0.03$ days, 
for data in the 1.7 hours to 1.2 day period\cite{Lee}.

%Moderski et al.~(2000), Huang at al. (2000a,b), 
%Kumar \& Panaitescu ( 2000) and 
%Panaitescu \& Kumar (2001) 

Various groups\cite{Moderski,Huang,Kumar,Pania} have modelled the
light emitted by firecones without some of the approximations originally
introduced by Rhoads. The evolution of the ejecta, for instance,
is treated continuously, not as a process with a break-time. 
%The emission is computed from isochronous 
%points in the firecone, so that light simultaneously received
%is light that had been simultaneously emitted
%(lifting the prior ``approximation'' that the speed of  light is infinite).
Not having an abrupt break put in by hand,
no abrupt break is predicted. The fair conclusion\cite{Moderski} is that
the AG light curves, {\it even in firecone models,} 
are too smooth to allow for a determination
of a hypothetical break time $t_b$. There is no reason to fit them
with underived expressions such as Eq.~(\ref{pheno}), nor
to extract any conclusions from such fits.

\begin{figure}
%\vspace*{-210pt}
\hspace{1.5cm}
%\leftline{\hfill\vbox{\hrule width 5cm height0.001pt}\hfill}
\epsfig{figure=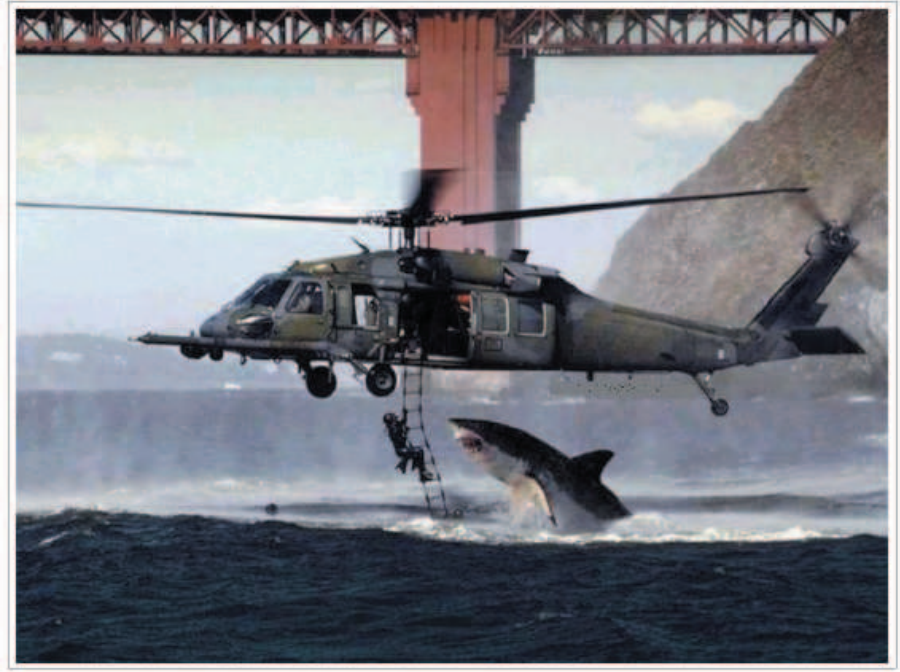,width=12.0cm}
\vspace*{-.truein}		
%\leftline{\hfill\vbox{\hrule width 5cm height0.001pt}\hfill}
\caption{The author trying to exit from a GRB conference.}
\label{fig:2}
\end{figure}

To me, the most surprising aspect of all detailed theoretical analyses of
GRB AG data published so far is that the firecone advocates
{\it  place the observer precisely on the jet's axis}, 
for no stated reason. It is obvious that the viewing
angle is a relevant parameter that cannot be unceremoniously dismissed,
if only because $d\cos\theta=\sin\theta\,d\theta$: the probability
of being away from the axis increases with $\theta$.
Moreover, a distribution of viewing angles would completely erase
a possible meaning of the distribution\cite{Frail}
 of specific $t_b$ values extracted
from expressions such as Eq.~(\ref{pheno}).
To say the least, the {\it anthropo-axial} view that the
ejecta of the observed AGs always point to the observer has not been 
shown to be a fair approximation. To air this kind of critique at
GRB conferences is not without danger, as made evident by
Fig.~\ref{fig:2}.

%The hypothetical firetrumpet ejecta behave in a
% different way from most of the highly relativistic jets observed
%in quasars (e.g. radio jets:
%Bridle 2000; optical jets: Cranc et al.~1993; X-ray jets: Wilson et al.~2000)
%and microquasars (e.g. Mirabel \& Rodriguez
%1994, 1999). The ejecta of the real jets, as seen from their emission
%point up to the point where they eventually stop and expand, generally
%subtend angles that {\it decrease} with time, exactly the opposite
%of the assumed firetrumpet behaviour of Eq.~(\ref{jetangle})
%and Fig.~(\ref{figtrumpet}a).
%In the analysis of these real objects (e.g. Pearsons \&
%Zensus 1987; Mirabel \& Rodriguez 1994, 1999; Ghisellini \& Celotti
%2001) it is the angle of observation ---and not the angle subtended by 
%the ejecta--- that plays a key role.

\subsection{The AG of GRB 991208 in fireball language}

The summary of this detailed section is: the SM ---in its many variations
and with lots of parameters--- fails in its fits and predictions and, to face this fact,
only unsupported excuses are given.

Early radio measurements of the AG of this GRB, from day $\sim\! 3$
 to day $\sim\! 14$ after burst, were reported by Galama et al.\cite{G1} 
(hereafter G1), who discussed them, along with the optical data, in the ``standard''
fireball model (hereafter SM). Follow-up radio measurements up to day 293 after 
burst and a SM reanalysis of the broad-band
AG of GRB 991208 were reported by Galama et al.\cite{G2} (hereafter G2). 

In G1 the authors use the radio and optical AG of this GRB
 to extract the power-law index of the
synchrotron-radiating electron distribution $p$, and the values
at three different times of the SM quantities needed to describe
the AG: the self-absorption and peak frequencies $\nu_a$ and
$\nu_m$, and the peak flux density $F_m$.  With $p=2.52$ fixed
and power-law fits to the temporal evolution of the other quantities,
they fit the AG of GRB 991208. They find that various spherical models
(``ISM" with a constant-density medium, and ``WIND''
with a circumburst ISM density decreasing as $1/r^2$) and a conical (``JET") 
model are all inadequate.  In conclusion,
they advocate a model with a transition from a quasi-spherical to
a jet evolution, but they do not offer 
any analysis in its support: the conclusion is what the French
would call {\it paroles verbales}.  The predictions of G1 are that $
\nu_a\propto t^{-14/13}$ at $ t\!>\! 10$ days and $F_\nu\propto
t^{-(2.2\; to \; 2.5)}$ for $\nu=8.46$ GHz at $ t\!>\! 12$ days,
as well as for $\nu=4.86$ GHz at $ t\!>\! 17$ days. These predictions are
in stark contrast with the findings of G2.

In G2  the authors introduce an extra cooling frequency
$\nu_c$, and a ``FREE'' fit with 9 parameters
($p$ and 2 parameters for $F_m$ and
each transition frequency, all assumed
to behave as $C\,t^{-b}$). The more constrained
ISM and WIND models turn out to be  
inadequate, as in G1.  The JET model is an
improvement over that of G1, but it fails to reconcile the late-time
decay $ F_\nu\sim t^{-1.1}$ at 8.46 GHz with the much steeper
optical decay\cite{sa00}  $ F_R\propto t^{-2.2\pm 0.1}$, which should be
similar. The FREE model provides a satisfactory fit to
the data, but it implies that the combination $\gamma\,B^3$ of the
bulk Lorentz factor of the flow and the post-shock magnetic field
ought to be roughly constant, while both are expected to decline
with time. In G2, the predictions\cite{li01} of
a model with two electron
energy distributions are also found to fail.

Faced with so much unsuccess, the authors of G2 conclude 
``the simplest
explanation which is consistent with the data and requires no
significant modifications is that the blast wave of GRB 991208
entered a non-relativistic expansion phase several months after
the burst''. As in G1, no support is given to this verbal 
conclusion.

\subsection{A refreshing interlude: Jets in Astrophysics}
  
\begin{figure}
\vspace*{-370pt}
\hspace{-.4cm}
%\leftline{\hfill\vbox{\hrule width 5cm height0.001pt}\hfill}
%\epsfig{figure=PictorA.eps_2.eps,width=14.0cm}
\epsfig{figure=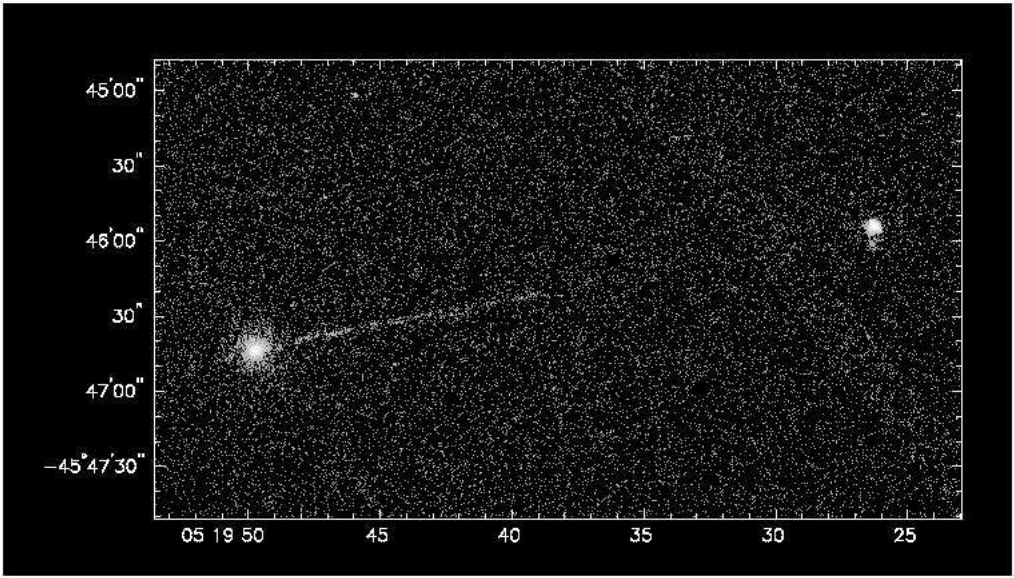,width=15cm}
\vspace*{-.4truein}		
%\leftline{\hfill\vbox{\hrule width 5cm height0.001pt}\hfill}
\caption{A Chandra X-Ray image of Pictor A.}
\label{fig:3}
\end{figure}

\begin{figure}
\vspace*{-0truecm}
\hspace{.5cm}
%\leftline{\hfill\vbox{\hrule width 5cm height0.001pt}\hfill}
\epsfig{figure=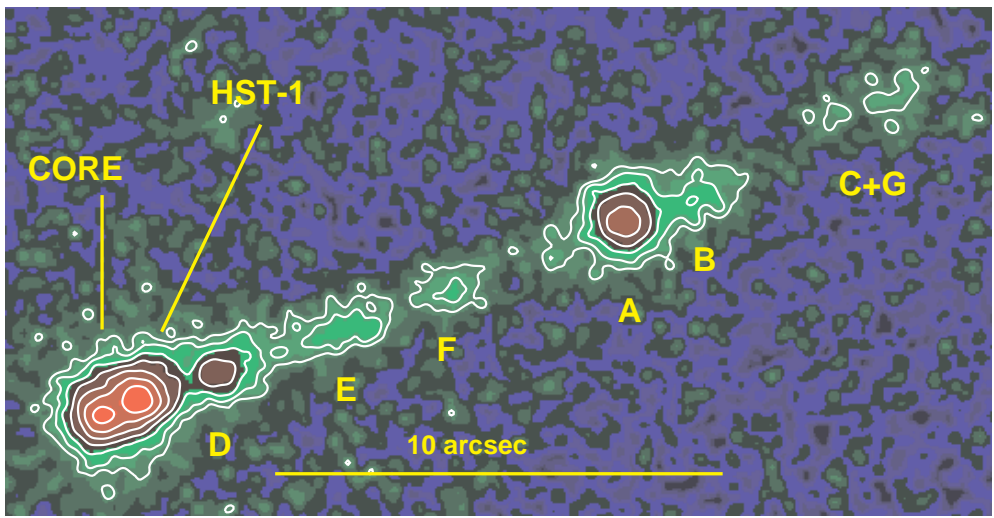,width=13.2cm}
%\vspace*{-1truein}		
%\leftline{\hfill\vbox{\hrule width 5cm height0.001pt}\hfill}
\caption{A Chandra X-Ray image of the M87 jet.}
\label{fig:4}
\end{figure}

A look at the sky, or a more modest one at the web, results in the
realization that jets are emitted by very many astrophysical systems
(stars, quasars, microquasars,...). One of the most impressive cases\cite{Wilson}
is that of the quasar Pictor A, shown in Fig.~\ref{fig:3}. Somehow, the active
galactic nucleus of this object is spitting something that does not
appear to expand sideways before it stops and blows up, having by then
travelled for a distance of several times the visible radius of a
galaxy such as ours. 
A closer look at another such object is provided
in Fig.~\ref{fig:4}: the central region 
of M87, wherein what appears
to be a succession of blobs of matter has been jetted\cite{Harris}. Many such 
systems have been observed. They are very relativistic: the Lorentz factors
$\gamma\equiv E/(mc^2)$ of their ejecta are typically of 
${\cal{O}}(10)$, and there is one case with a claimed\cite{KC} $\gamma\sim 10^3$.
The mechanisms responsible for these mighty ejections ---suspected
to be due to episodes of violent accretion into a very
massive black hole--- is not understood.

\begin{figure}
\vspace*{-8cm}
\vspace{10pt}
%\leftline{\hfill\vbox{\hrule width 5cm height0.001pt}\hfill}
\epsfig{figure=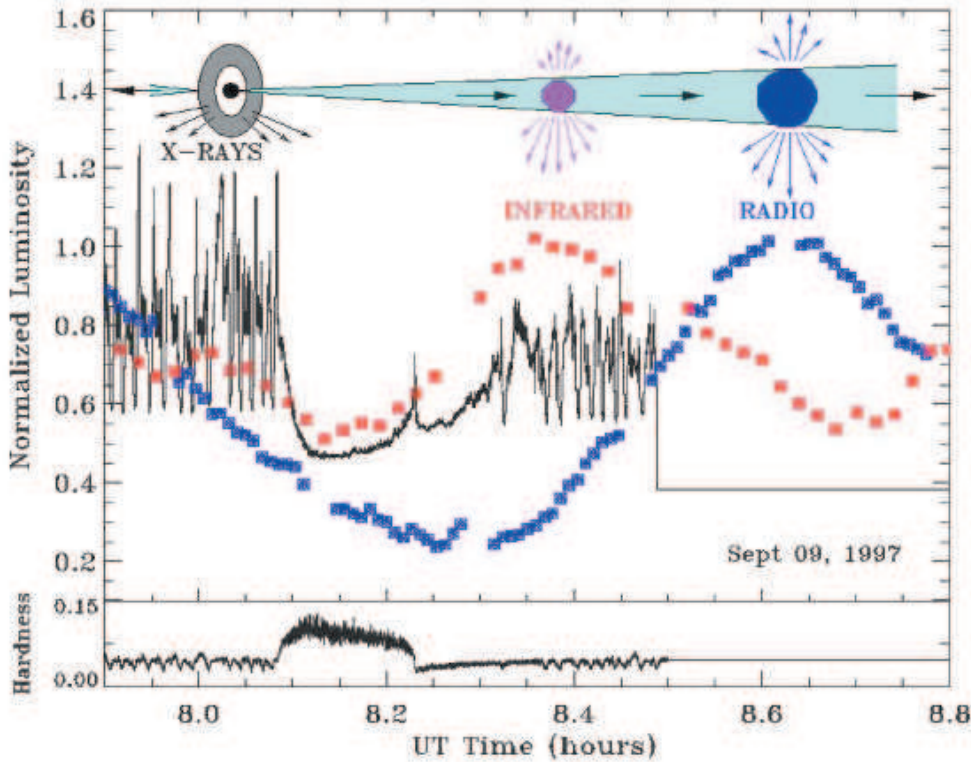,width=14.0cm}
\vspace*{-1truein}		
%\leftline{\hfill\vbox{\hrule width 5cm height0.001pt}\hfill}
\caption{X-ray, infrared and radio fluence of GRS 1915+105, as it emits
two oppositely flying cannonballs.}
\label{fig:5}
\end{figure}

In our galaxy there are ``micro-quasars'', in which the central black
hole is only a few times more massive than the Sun. The best studied 
example\cite{Felix} is the $\gamma$-ray source  GRS 1915+105.
In a non-periodic manner, about once a month, this object emits two
oppositely directed {\it cannonballs}, travelling at $v\sim 0.92\, c$.
As the emission takes place, and as illustrated in Fig.~\ref{fig:5}, the X-ray 
emission ---attributed to an unstable accretion disk--- temporarily decreases.
How part of the accreting material ends up ejected along the system's axis
is not understood. The process reminds one of the blobs emitted
upwards as the water closes into the ``hole'' made by a stone
dropped onto its surface. It is only the relativistic, general-relativistic
magneto-hydro-dynamic details that remain to be filled in!
Atomic lines from many elements have been observed\cite{Kotani} in
the CBs of $\mu$-quasar SS 433. Thus, at least in this case, the
ejecta are made of ordinary matter, and not of some fancier substance
such as $e^+e^-$ pairs. In the analysis of quasar and microquasar
ejecta, the relevant parameters are the Lorentz factor and the angle
between the jet direction and the observer (and not, as in the firecone
models, the opening angle of the jetted material).

%\vskip-.5cm

\subsection{The Cannonball model}

The CB model is not based on hypothetical fireballs, firecones 
or {\it hypernovae}, but on close
analogies with the observed astrophysical
jets. The  {\it long-duration} GRBs and 
their AGs are produced in ordinary {\it core-collapse} supernovae by jets
of CBs, made of {\it ordinary atomic matter}, and travelling  with  high Lorentz
factors, $\gamma\sim 10^3$.  A CB is emitted, as observed in
$\mu$-quasars, when part of an accretion disk falls abruptly onto
the newly-born compact object. As it crosses the circumburst material
(the shell of the SN and the ``wind'' of the parent star)
the surface of a CB is collisionally heated to keV temperatures
and the quasi-thermal radiation it emits as it reaches the transparent
outskirts of the circumburst matter ---boosted and collimated by the CB's
motion--- is a {\it single $\gamma$-ray pulse} in a GRB.
A competing mechanism\cite{DS,dd0} for the $\gamma$-ray production
is inverse Compton scattering of the
SN light by the electrons in the emerging CBs\footnote{This process produces
a highly polarized beam of $\gamma$-rays, except right at the jet's axis.}.
For both mechanisms, the timing sequence of emission of the successive 
individual pulses (or CBs) in a GRB reflects the chaotic accretion process;
its properties are not predictable, but those of the single pulses are.

After a few observer minutes, the CBs' emissivity --the afterglow--
is dominated by synchrotron emission from the electrons that
impinge and penetrate in them from the ISM\cite{ddd2a}.  
These electrons
are Fermi-accelerated in the CB's magnetic maze to a broken power-law
energy distribution with a ``bend'' energy $E_b=\gamma(t)\,m_e\,c^2$,
with $\gamma(t)$ diminishing as the CBs decelerate.  Let $\theta$
be the angle between the CBs' viewing and travelling directions,
$\delta(t)\approx 2\,\gamma/(1+\gamma^2\theta^2)$ be the emitted
radiation's Doppler energy-boost factor, and $n_e\approx n_p$ the electron 
and proton ISM number densities. The observer sees an energy flux:
\begin{equation}
F_\nu \equiv \nu\, {dn_\gamma\over d\,\nu} \propto n_e\, 
[\gamma(t)]^{3\alpha-1}\, [\delta(t)]^{3+\alpha}\, \nu^{-\alpha}\, , 
\label{sync}
\end{equation}
with $ \alpha$ steepening from $\approx 0.5$ to
$\approx p/2 \approx 1.1$ at the ``injection bend'' frequency\cite{ddd2b}
corresponding to $ E_b $:
\begin{equation}
\nu_b \simeq {1.87\times 10^3\, [\gamma(t)]^3\, \delta(t)\over 1+z}\, 
\left [{n_p\over 10^{-3}\,{\rm cm}^{-3}}\right]^{1/2}\, {\rm Hz}\, .
\label{nubend}
\end{equation}

We assume the SNe associated with GRBs to occur mainly
in super-bubbles of approximately constant $n_p$, except
in the immediate vicinity of the progenitor.
For a constant-density ISM, $\gamma(t)$
is the real root of the cubic:
\begin{equation}
{1\over\gamma^3}-{1\over\gamma_0^3}
+3\,\theta^2\,\left[{1\over\gamma}-{1\over\gamma_0}\right]=
{6\,c\, t\over (1+z)\, x_\infty}\; ,
\label{cubic}
\end{equation}
where $\gamma_0=\gamma(0)$, and
$ x_\infty$
characterizes the CB's slow-down (it takes a distance $x_\infty/\gamma_0$
for a CB to decelerate to half its original Lorentz factor).

In the CB-model fits\cite{ddd2a} we fix the electron index to its
theoretical value $p\approx 2.2$.  The optical AGs then depend on
4 parameters:  overall normalization, $\gamma_0$, $x_\infty$ and
$\theta$. In the radio domain, where self-absorption is important,
only one extra parameter is needed.  The dominant absorption
mechanism is free--free attenuation (inverse bremsstrahlung
$e\,p\,\gamma\rightarrow e\,p$), characterized by a parameter
$\nu_a$ in the opacity, which behaves as $\tau_\nu=(\nu_a/\nu)^2\,
(\gamma(t)/\gamma_0)^2$.  Self-absorption produces  a turn-around
of the spectra from $ F_\nu\sim \nu^{1.5}$ to  $ F_\nu\sim \nu^{-0.5}$
as $\nu$ increases.  

To date, the CB model provides a very good, simple and unified description 
of the 
light curves and spectra of the AGs of {\it all} GRBs of known redshift. 

%The description of a radio AG involves two
%additional effects which, in fair approximations, introduce no
%extra parameters:  a ``cumulation factor'' describing the time it
%takes the ISM electrons gathered by the CB to cool to the lower
%radio-emitting energies, and an ``illumination and limb-darkening
%factor'' taking into account that the CBs are viewed relativistically
%(an observer would ``see'' almost all of the $4\pi$ surface of a
%spherical CB).  With these corrections, Eq.~(\ref{sync}) provides
%good fits to the broad-band AGs of all GRBs of known redshift
%(DDD2).

\subsection{GRB 991208 in the CB model}

This section is to the CB model what section 3.3 is to the 
standard model.
Its summary is: the CB model ---simple and
parameter-thrifty as it is--- succeeds in its fits and predictions.

The AG of GRB 991208 has
been analysed thrice in the realm of the CB model. In
Dado et al.\cite{ddd2a} (hereafter DDD1) we fitted the available 
R-band data on all the then-measured GRBs of known redshift, 
including this one, which is shown in the left panel of Fig.~\ref{fig:1}. In 
DDD2\cite{ddd2b} these results were extended 
to wide-band fits of all the available optical and radio data, which included at
the time the early radio data of G1\cite{G1}. Finally, in DDD3\cite{ddd3}, we 
compared our predictions in DDD2 with the observations in G2\cite{G2} 
of the later-time radio data.  

In the CB model, the AG of a GRB has three contributions: the CB, the host
galaxy and a ``standard candle'' supernova akin to SN1998bw,
transported to the GRB's redshift\footnote{In the CB model, GRB
980425 ---associated to SN1998bw--- is in no way 
exceptional\cite{dd0,ddd2a,ddd2b}. Unlike in the SM, it makes sense
to use this SN as a putative standard candle.}.  In
Fig.~{\ref{fig:1} the presence of the SN is shown in two cases,
one of which is GRB 991208. In
a CB-model analysis, in all instances wherein such a SN could be
seen, it was seen. This is so for all AGs with $ z\!<\!1.2$
(DDD1), including the cases where the presence of a 1998bw-like SN
was a prediction, based on the optical data {\it preceding} the
observable SN contribution\cite{ddd2c,ddd2d,ddd3c,ourSN}.

Results for multifrequency fits to GRB AGs cannot be shown in a limited
space. In comparing the SM results of G1 and G2 with
the CB-model results of DDD2 and DDD3, I concentrate
on the radio data at 8.46 GHz, for which the measurements of G1 and G2
are particularly abundant and precise.
Shown in Fig.~\ref{fig:6} are, on the left, the G1 early data and its SM 
extrapolated prediction for later times, quoted in section 3.3. Shown on the right
are the CB model fit of DDD2 and its predicted extrapolation in time.

\begin{figure}
\vspace*{-10pt}
%\leftline{\hfill\vbox{\hrule width 5cm height0.001pt}\hfill}
\hspace{.3cm}
\mbox{
\epsfig{figure=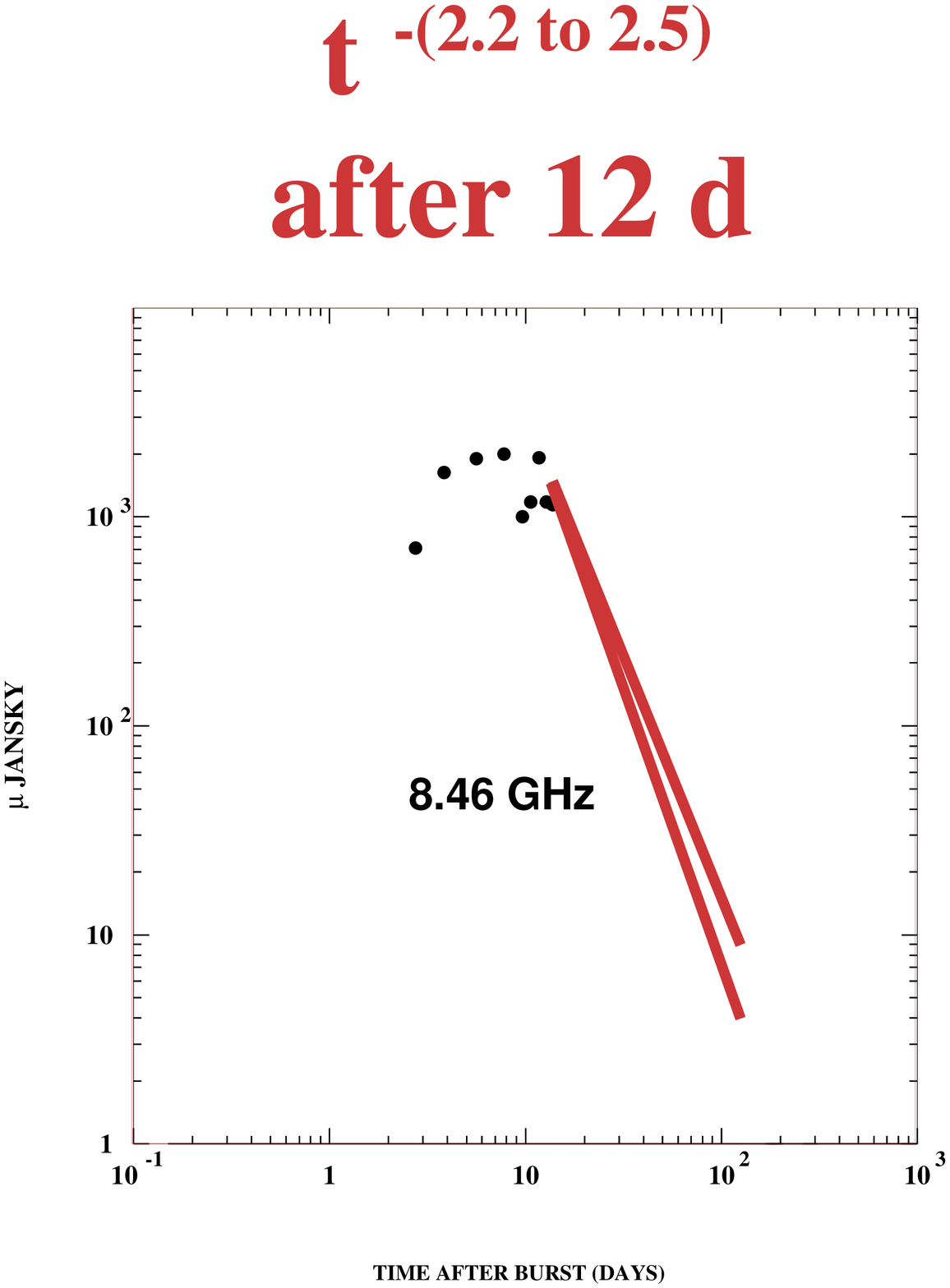,width=7.cm}
\epsfig{figure=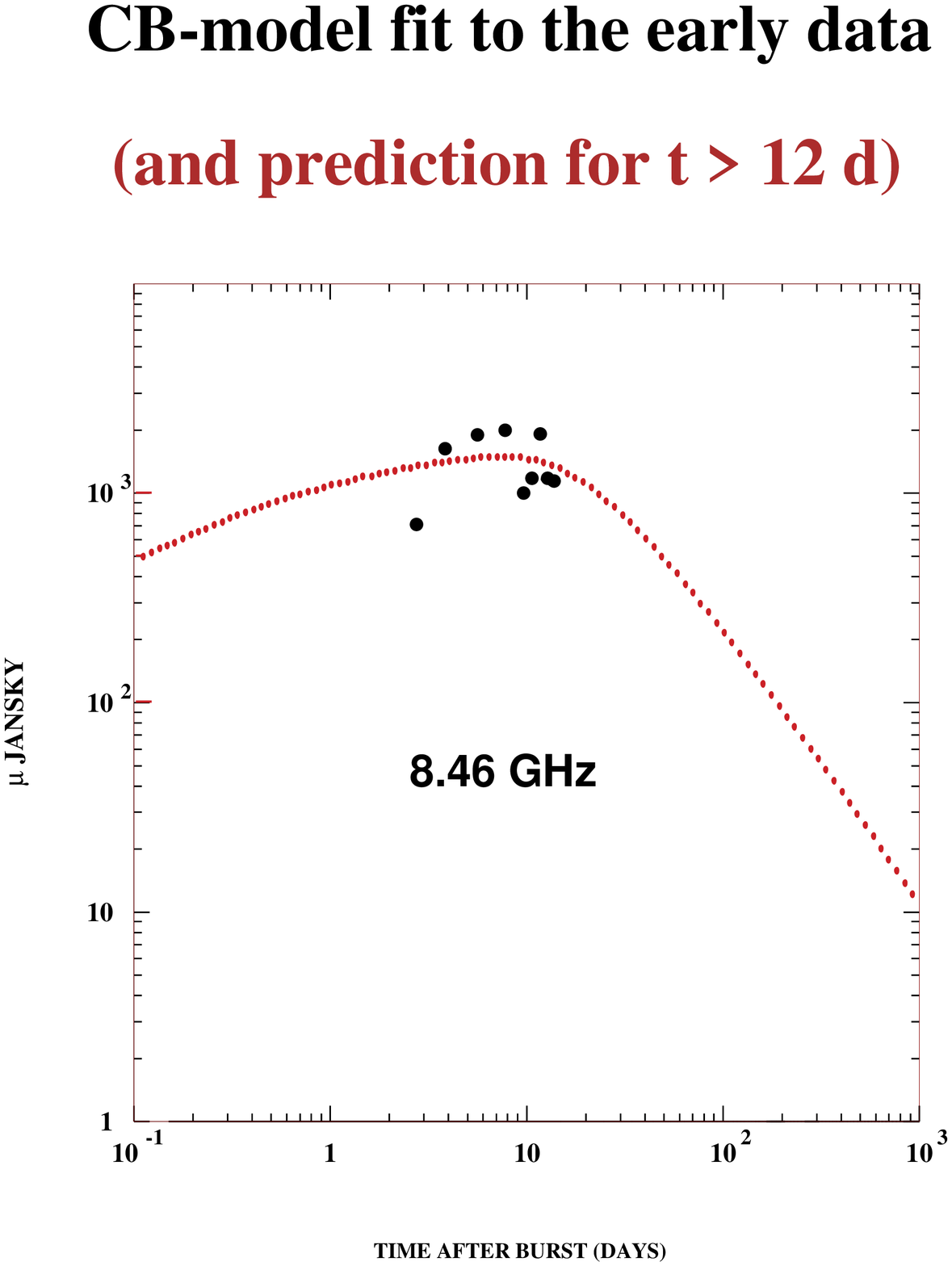,width=7.cm}}
%\epsfig{figure=123Rbandnew.eps,width=7.5cm}}
%\vspace*{-0.4truein}		
%\leftline{\hfill\vbox{\hrule width 5cm height0.001pt}\hfill}
\caption{The radio fluence of GRB 991208 at 8.46 GHz (G1). The ups and downs
of the data are due to scintillations. The best fits 
shown are the subensemble of the wide-band fits by G1 (left) and DDD2 (right)
that refer to this particular radio-frequency. The (red) thick lines in
the left panel bracket the predictions of G1.}
\label{fig:6}
\end{figure} 

Referring once again to an observed 8.46 GHz frequency, I show in
Fig.~\ref{fig:7}, on the left, the comparison of the late
G2 radio data with the SM prediction in G1. The later data are
labelled by asterisks. In the same figure,
on its right, two CB-model fits are shown. One of these wide-band
fits is the DDD2 fit shown in Fig.~\ref{fig:6}, the other is a new fit,
along identical lines, including the new radio data of G2.
The figure shows that the predictions of DDD2 were very satisfactory:
the a-priori and a-posteriori fits are very similar and they both provide a  
good description of the observations, their difference not being larger
than the scintillating ups and downs of the data.

\begin{figure}
\vspace*{-10pt}
%\leftline{\hfill\vbox{\hrule width 5cm height0.001pt}\hfill}
\hspace{.3cm}
\mbox{
\epsfig{figure=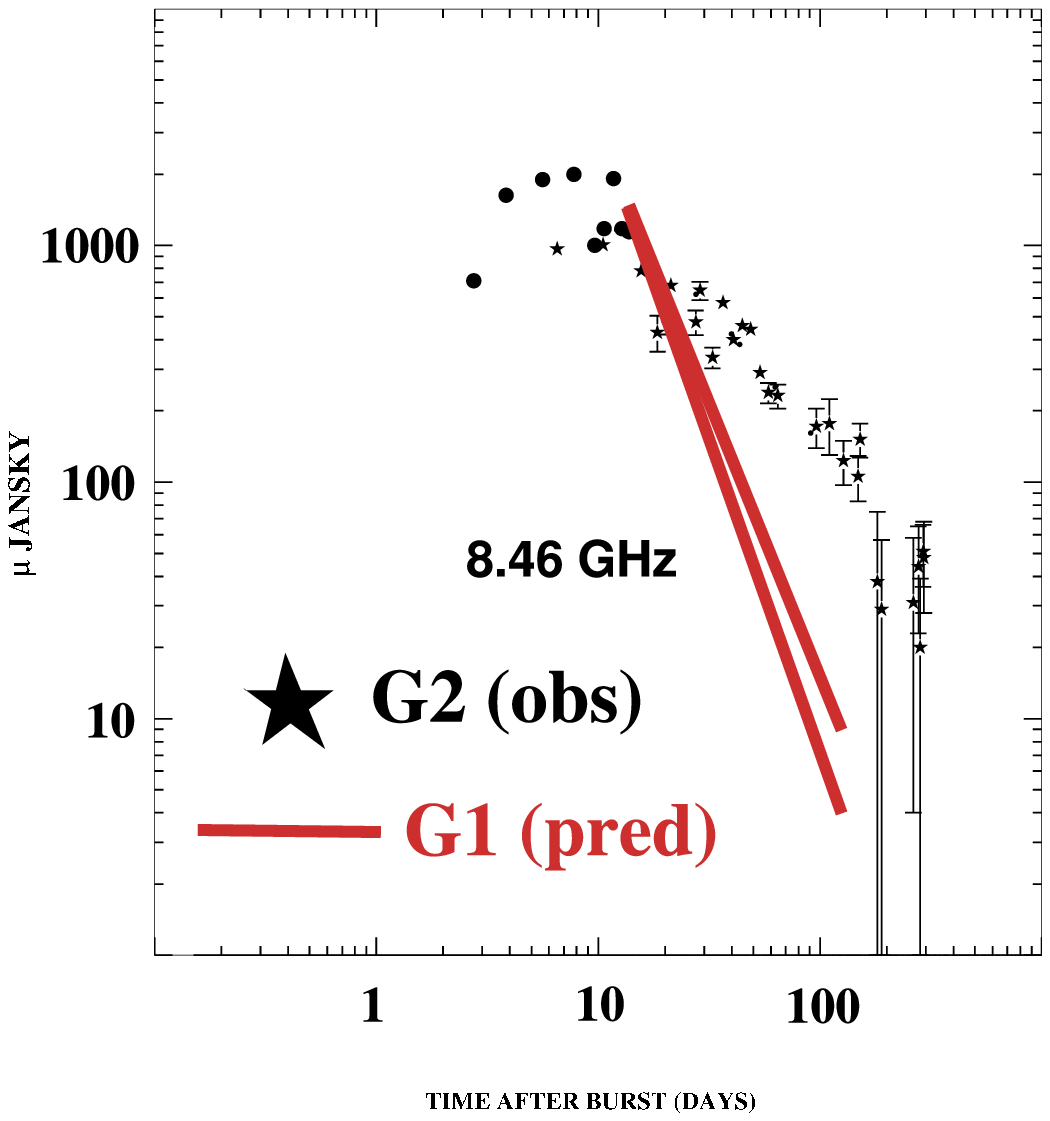,width=6.3cm}  
\epsfig{figure=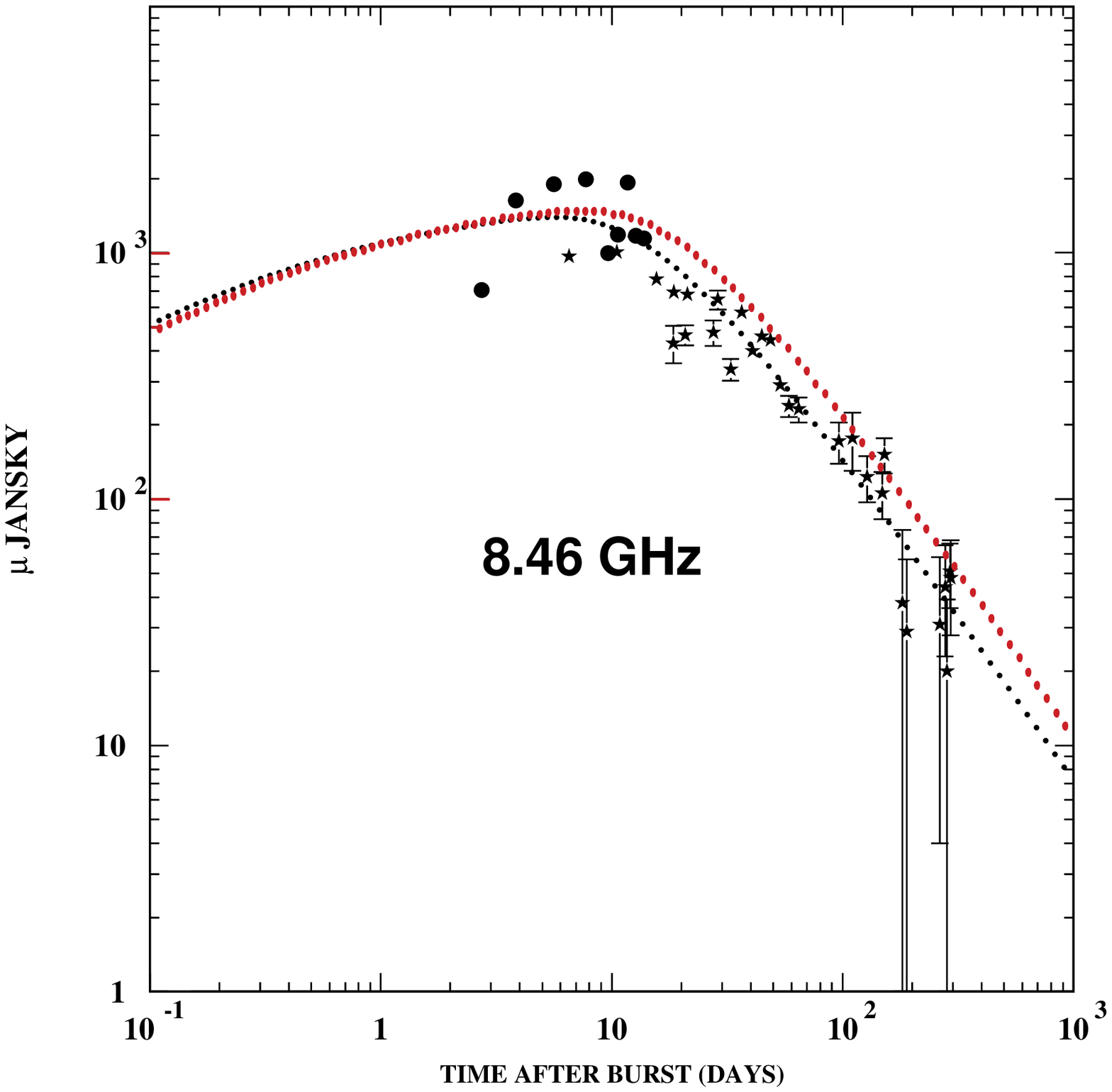,width=7.4cm}}
%\vspace*{-0.4truein}		
%\leftline{\hfill\vbox{\hrule width 5cm height0.001pt}\hfill}
\caption{Comparisons of the predictions of G1 (left) and DDD2 (right)
with the later measurements in G2, labelled by asterisks. The red-point line
on the right is the CB-model prediction of DDD2 and Fig.~\ref{fig:6}. 
The black-point line
is a fit including the G2 data.}
\label{fig:7}
\end{figure}

In Table 1 I report the parameters for the R-band fit of DDD1,
the broad-band fit of DDD2 and the DDD3 fit to all data. They are
quite stable. Even the DDD1 fit to only the R-band data determines
$\gamma_0$, $\theta$ and $ x_\infty$ to within a few per-cent of
the results of the DDD2 fit (117 data points in total), even though
the former fit is based on the mere dozen of early data points that
are not dominated by the SN, see Fig.~\ref{fig:1}.
%The value of $ x_\infty$ in the DDD2
%fit is a bit smaller than in the others, the reason being that
%---as can be seen by inspection of Fig. \ref{fig:7}, ---
%this parameter is sensitive to the late observations, and the early
%radio data of G1 dominated the DDD2 fit. 
For fits that are so similar,
their single parameters describing the overall normalization are
also necessarily similar:  they are not reported in Table 1.

%\vskip 1.0cm
%\noindent
%{\bf Table I:} Successive CB-model fits to the AG of GRB 991208.
%R-band is a fit to only that optical frequency (DDD1). WB1 is the
%wide-band fit in DDD2, with only the early radio-data. WB2 is
%the current fit to all data.}
%%\vskip 0.3 true cm
%\begin{table}[h]
%%\huge\bf
%\normalsize
%\hspace{3.0cm} %if you want to center your table act on this argument
%\begin{tabular}{|l|c|c|c|c|}
%\hline
%Parameter & R-band & WB1 & WB2 \\
%\hline
%$\theta$ [mrad]           & 0.100 & 0.111    & 0.103  \\
%$\gamma_0$            & 1034  &  1034    & 1089   \\
%$ x_{\infty} $ [Mpc]            & 1.357 & 1.014    & 1.382  \\
%$\nu_a $ [MHz]           & ****      &   103    &    98  \\
%\hline
%%
%\end{tabular}
%\end{table}

\begin{table}[h]
\caption{Parameters of the successive CB-model fits to the AG of GRB 991208.
DDD1 is a fit to only R-band optical frequencies. DDD2 is a
wide-band fit with only the early radio-data of G1. DDD2 is
the fit to all data, including the G2 late radio data.}
\label{tab:par}
  \small
  \vspace{.5cm}
  \hspace{4.0cm}
  \begin{tabular}{||c|c|c|l||}\hline\hline
  {} &{} &{} &{}\\
  Parameter &DDD1 &DDD2 &DDD3\\
  {} &{} &{} &{}\\
  \hline\hline
  {} &{} &{} &{}\\
  $\theta$ [mrad]           & 0.100 & 0.111    & 0.103  \\
  {} &{} &{} &{}\\
  \hline
  {} &{} &{} &{}\\
  $\gamma_0$            & 1034  &  1034    & 1089   \\ 
  {} &{} &{} &{}\\
  \hline
  {} &{} &{} &{}\\
  $ x_{\infty} $ [Mpc]            & 1.357 & 1.014    & 1.382  \\
  {} &{} &{} &{}\\
  \hline
  {} &{} &{} &{}\\
  $\nu_a $ [MHz]           & ****      &   103    &    98  \\
  {} &{} &{} &{}\\
\hline\hline
\end{tabular} 
\end{table}

Table 1 shows how robust and predictive
the CB model is. Its parameters
are few, and consistently determined as new data are added to successive fits.
This, as we have seen, is in stark contrast with the situation in the SM, with 
its many more parameters,
its predictive failures, and the verbal excuses for its successive inadequacies.
In my opinion Occam would have no trouble choosing between these models;
he may not even need his proverbial sharp razor, a dull spoon would suffice. 
Notice that, the way I put it, the defenders of the standard model would also agree!

\section{Sociology}

In what follows, I  try to place the avatars of the CB model
in a more general context.

\subsection{Things that happen to us}

So far, the CB model has encountered no problems in its confrontation
with data. It is in its confrontation with humans that the model fares the worst.
Just as an example, let me comment on the result of having sent for publication
in the Astrophysical Journal Letters our paper
DDD3, entitled ``Fireballs and Cannonballs confront GRB 991208'', where
we compare the SM analysis by G1 and G2 to our CB-model analysis of the
same data, as I did here. 

The first referee asserts: {\it ``There is nothing new in this manuscript''}.
Then, commenting on the fact that Galama et al.~do not refer to our
successful predictions of their G2 data, he\footnote{I am not being sexist
in my language, I am sure that no woman would have said such things.}
continues, somewhat contradicting himself: {\it ``If Dado et al.~are right,
then surely the referee of Galama et al.~will catch this gross oversight
and have it fixed in the revised submission".} This is pulling our 
legs under the cover of anonymity.

We ask for a second referee, who reports: {\it ``The claim that the CB model
works better than the standard fireball models is not made in a way that
would convince any objective reader".} He does not bother to say why.
To me this sounds as being impermeable to facts.
Finally, we complain to the editor, who responds: {\it ``I have no claim to
infallibility".} 

My colleagues and I have written some 18 papers on GRBs in the CB model,
the first four of which were rejected on grounds very similar to the above.
That is, {\it not once} on the basis of scientific critique. This is very discouraging.
One is tempted to entitle all of one's papers on the subject  ``Mission Impossible".

\subsection{Things that happen}

Scientists generally adhere to the Popperian\cite{Popper} view that 
{\it every genuine test of a theory is an attempt to falsify it.} Reading the
current astrophysical literature, one very often sees this opinion challenged.
As we have seen, and to quote Popper again:
{\it Some genuinely testable theories, when found to be false, are still upheld by their 
admirers ---for example by introducing ad hoc some auxiliary assumption, 
or by reinterpreting the theory ad hoc in such a way that it escapes refutation. Such a 
procedure is always possible, but it rescues the theory from refutation only at 
the price of destroying, or at least lowering, its scientific status.} But it can be 
even worse, as I shall illustrate with a few examples.

GRBs occur in star-formation regions and are no doubt associated
with processes in aged very large stars. One expects the matter density around
the progenitor to be characteristic of the environment of such stars,
which shed matter for a few thousand years at the end of their lives,
at a roughly constant mass-loss rate. This results in a 
circumstellar density profile $\rho\!\sim\! 1/r^2$.
In an article\cite{Price} on GRB 011121, its authors state: {\it ``Unfortunately,
until now there has been no clear evidence for a wind-fed circumburst medium...
in the afterglow of any cosmologically located GRB''}. At the time there were
a score of such AGs, meaning that {\it unfortunately} the SM expectation was 
not corroborated for these many cases. The article continues: {\it ``We undertake
afterglow modeling of this important event and to our delight have found 
a good case for a wind-fed circumburst medium".} This kind of 
{\bf misfortune} and {\bf delight}
may be what Popper refers to when saying: {\it It is easy to obtain confirmations, or verifications, for nearly every theory ---if we look for confirmations.} {\bf Fortunately,}
this attitude is the opposite to that of most researchers in most areas of science,
who are motivated by challenging their respective ``standard models''.

Placing belief in prejudices ahead of facts that contradict them is not as 
unusual in science as one may hope. Cosmic-ray physics is another
area in which the corresponding ``standard model'' is in dire straights.
Cosmic rays are supposed to be accelerated in the shocks produced
by the collisions of the moving
shells resulting from SN explosions with the ambient matter.
Synchrotron radiation from accelerated electrons in such sites has been
observed, but the limits on $\gamma$ rays from $\pi^0$ decay are well
below the expectation from collisions of the assumed accelerated protons 
and nuclei.
I looked for a recent review on the subject and, in the first one I found\cite{GL},
it says: {\bf ``Frustratingly,} {\it the accelerated ``cosmic-ray'' ions, which should 
produce
$\gamma$ rays by nuclear interactions with the ambient gas, have evaded detection
so far. So, we are still unable} {\bf to prove} {\it the origin of cosmic rays
nearly a century after their discovery!''.} The emphasis is mine, the style, by now,
is familiar.

%\subsection{Things that happen in other areas}

More often than not, astrophysical phenomena involve very complex
physics. Deciphering them is not unlike deciphering the script of an
ancient language. It may not be surprising that there are similarities between
the sociologies of these two fields.

The jesuit priest Athanasius Kircher pretended to have deciphered
the ancient-Egyptian hieroglyphic language. His ``translations'' of
the writings in the Egyptian obelisks in Rome
were entirely forged. But Kircher's influence
was so enormous that he managed to stall progress in the field for
a full century. Alas, even in the XXth century, similar things did happen.
Sir John Eric Sidney Thompson had an entirely incorrect view of 
Mayan scriptures, which he thought were ``anagogical''
(having a mystical meaning beyond the text) and ideographic, as
opposed to phonetic. Given his great stature in the field, his
views delayed the deciphering of the Mayan language by some 30 years. 
Perhaps he was knighted for that feat.

Nothing of what I have reported is {\bf that} surprising for, after
all, science is a human endeavour, not a divine activity. Yet,
by now, if I am being effective, my audience or my hypothetical readers
ought to be indignant. {\it Nothing of this sort could happen in my
university or my lab!}... should they be screaming to themselves. I
cannot fully scream in the same way for, in my lab (though it is
the ``best and biggest'' of its kind) one can also find all sorts of
humans. Which brings me to my next subject.

\subsection{High Energy Physics}

Two things that high-energy physicists cannot be accused of
are excessive belief in their standard model (in spite of its brutal
and continued success) and lack of receptiveness to new ideas.
A result lying three or even fewer standard deviations away
from the standard expectation
 is enough to send a good fraction of the community
bananas. Theories that have very little contact with current
experiments are widely venerated. Models that would require
an extraordinary piece of luck to be relevant in the near future
(such as extra dimensions at accessible high energies or short 
distances) create a flurry of activity. None of this is particularly
unhealthy, for science is exploration by the open-minded.

Yet, the fact that HEP experimental research now requires very large
groups and a complex organization inevitably brings into the fore
considerations that are not as ``pure'' as ``pure science'' ought to be,
according to ``the book". Thus, we announce with great to-do
``discoveries''
that are not all that convincing. We embark in programs that
are patently useless. When giving the talks on which this article
is based to HEP audiences, I did not say explicitly to what 
discoveries or programs I
was referring to. And yet... everybody knew.
Exactly.

In my opinion, the most worrisome current aspect of HEP in not
that we are moving so fast from science to management,
or from science to ``post-academic'' science. What really bothers me 
is how little we are doing to revitalize our field by fighting for more
diversified scientific programs, for larger long-term investments
in R\&D, and for new resources, even though the field is
as challenging as it ever was!

\section{Conclusions}

It goes without saying that my critique of the standard model of GRBs
does not imply that I am convinced that the cannonball model is 
entirely ``right''.
It is an extreme simplification of a very complicated phenomenon;
it will either require refinements or even turn out to be completely
wrong. The latter is improbable for, unlike the SM, the CB
model explains well and in an
extremely simple fashion practically all aspects of the data
on long-duration GRBs. There seems to be
something right about it. But the assumptions and predictions 
of the CB model, or any other, should 
be tested against
observations, or challenged for consistency. In other realms of
science the existence of a sensible model challenging the standard
lore would be very welcome, as opposed to olympically ignored.

Relative to my more general ``sociological'' considerations, I have nothing
to add. 
  
\section{Acknowledgements}
Viki Weisskopf used to say {\it ``Physics is best done in a hostile environment''.}
Surely my coauthors and I are, in this sense, obliged to quite a few ominous
and anonymous colleagues. But we would not have been able to survive the
environment for long, had we not held prior tenured positions.

 \end{document}